\documentclass[%
 reprint,
superscriptaddress,
nofootinbib,
longbibliography,
 amsmath,amssymb,aps,
 prx,
]{revtex4-2}

\usepackage{graphicx}
\usepackage{dcolumn}
\usepackage{bm}
\usepackage[hidelinks]{hyperref}

\usepackage{braket}

\PassOptionsToPackage{dvipsnames}{xcolor}

\usepackage{graphicx}
\usepackage{epsfig}
\usepackage{latexsym}
\usepackage{mathtools}
\usepackage{hyperref}
\usepackage{tikz-cd}
\usepackage[vcentermath]{youngtab}
\usepackage{mathrsfs}
\usepackage{url}
\usepackage{comment}\usetikzlibrary{decorations.markings,calc,intersections,through,backgrounds,shapes,snakes,shapes.geometric,knots,arrows,arrows.meta}
\usepackage[dvipsnames]{xcolor}
    
\tikzcdset{arrow style=tikz, diagrams={>=stealth}}

\def \be  {\begin{equation}}
\def \ee  {\end{equation}}
\def \bea {\begin{equation}\begin{aligned}}
\def \eea {\end{aligned}\end{equation}}
\def \ba  {\begin{eqnarray}}
\def \ea  {\end{eqnarray}}
\def \bb  {\begin{thebibliography}}
\def \eb  {\end{thebibliography}}
\def \lab #1 {\label{#1}}

\newcommand\cC{\mathcal{C}}

\newcommand\cI{\mathcal{I}}

\newcommand\cM{\mathcal{M}}

\newcommand\bC{\mathbb{C}}

\newcommand\bR{\mathbb{R}}

\newcommand\bZ{\mathbb{Z}}

\newcommand\Obj{\mathrm{Obj}}
\newcommand\Hom{\mathrm{Hom}}

\newcommand\vect{\mathbf{Vec}}

\newcommand\rep{\mathbf{Rep}}

\newcommand\str{\mathbf{Str}}

\definecolor{cardinal}{rgb}{0.6,0,0}
\definecolor{darkgreen}{rgb}{0,0.5,0}
\definecolor{golden}{rgb}{0.92, 0.7, 0}
\definecolor{midnight}{rgb}{0, 0, 0.5}
\definecolor{darkblue}{rgb}{0.2, 0, 0.8}

\tikzset{->-/.style={decoration={
  markings,
  mark=at position .5 with {\arrow{>}}},postaction={decorate}}}

\begin{document}

\title{Higher Representations and Quark Confinement}

\author{Finn Gagliano}
\email{finn.gagliano@durham.ac.uk}
\affiliation{%
 Department of Mathematical Sciences, Durham University, UK
}%
\author{Andrea Grigoletto}
\email{andrea.grigoletto@durham.ac.uk}
\affiliation{%
 Department of Mathematical Sciences, Durham University, UK
}%
\author{Kantaro Ohmori}
\email{kantaro@hep-th.phys.s.u-tokyo.ac.jp}
\affiliation{%
 Faculty of Science,
 University of Tokyo, Japan
}%

\begin{abstract}
The concept of a (de)confined phase in QFT is well-defined in the presence of $1$-form symmetries and their spontaneous symmetry breaking. However, in scenarios where such symmetries are absent, confinement is not a well-defined phase property. 
In this work, we propose that, when restricting to a specific submanifold of the parameter space -- namely at zero temperature and fixed quark mass -- the confined and adjoint Higgs phases of scalar QCD can be distinguished through the different organization of their spectra, as seen from the perspective of the baryon symmetry. 
The analysis is performed in terms of an appropriate higher-categorical representation theory, recently developed for generalized symmetries.
Consistent with expectations, we find that the confined phase permits only particles with integer baryon charges, while the Higgs phase is characterized by the coexistence of bare quarks and center vortices, exhibiting a non-trivial Aharonov-Bohm effect between these excitations.

\end{abstract}

\maketitle


\section{Introduction}

Symmetry is a cornerstone of quantum field theory, and the introduction of generalized global symmetries has expanded this concept in profound ways \cite{Gaiotto:2014kfa}. 
Since their inception, these symmetries have deepened our understanding of known phenomena, unveiled new structures in quantum systems, and advanced insights into e.g. selection rules \  \cite{Chang:2018iay,Huang:2021zvu, Robbins:2021ibx,Inamura:2021szw,Choi:2021kmx,Kaidi:2021xfk,Choi:2022zal,Li:2023ani,Bhardwaj:2023idu,Bhardwaj:2023fca,Bhardwaj:2024qrf} and associated generalized charges \cite{Chang:2018iay,Bullivant:2019fmk, Thorngren:2019iar,Thorngren:2021yso,Bartsch:2022mpm,Delcamp:2022sjf,Bartsch:2022ytj,Lin:2023uvm,Bartsch:2023pzl,Bhardwaj:2023wzd,Bhardwaj:2023ayw,Bartsch:2023wvv,Rayhaun:2023pgc,Cordova:2023qei,Copetti:2024rqj,Copetti:2024onh,Cordova:2024vsq,Cordova:2024iti,Copetti:2024dcz,Inamura:2024jke,Bhardwaj:2024igy,Choi:2024tri,Choi:2024wfm,Cordova:2024nux}.

Another notable area where generalized symmetry has made significant contributions is in understanding the phases of non-abelian gauge theories in $(3+1)$D
 \cite{Gaiotto:2017yup,Yamazaki:2017ulc, Tanizaki:2017bam, Shimizu:2017asf, Tanizaki:2017mtm, Anber:2018iof, Cordova:2018acb, Anber:2018jdf, Anber:2018xek, Hsin:2018vcg, Komargodski:2018odf, Wan:2018zql, Wan:2018djl, Yonekura:2019vyz, Hidaka:2019jtv, Wan:2019oyr, Karasik:2019bxn, Poppitz:2019fnp, Bolognesi:2019fej, Anber:2019nze, Cordova:2019bsd, Wang:2019obe, Tanizaki:2019rbk, Wan:2019oax, Karasik:2020pwu, Bolognesi:2020mpe, Sulejmanpasic:2020zfs, Chen:2020syd, Bub:2020mff, Anber:2021iip, Tanizaki:2022ngt, Morikawa:2022liz, Nguyen:2023fun, Hayashi:2023wwi, Dumitrescu:2023hbe, Chung:2024hsq, Hayashi:2024qkm, Hayashi:2024gxv, DelZotto:2024arv, Bolognesi:2024wje}. Historically, the correlation functions of Wilson loops have been used to characterize the phases of these theories via the identification
\begin{equation}
\langle W(\gamma) \rangle \sim
\begin{cases}
e^{-A(\gamma)} & \text{Confined phase}, \\
e^{-P(\gamma)} & \text{Deconfined phase},
\end{cases}
\label{eq:wilson-vev}
\end{equation}
where $A(\gamma)$ and $P(\gamma)$ are proportional to the minimal area and perimeter of the loop $\gamma$ \cite{Wilson:1974sk,tHooft:1977nqb}. For instance, these characterize respectively the confining phase of pure $SU(N)$ gauge theory and the Coulomb phase of pure $U(1)$ gauge theory.

Phrasing this behavior in the language of generalized symmetries gives rise to an extended Landau paradigm, which classifies phases based on whether these are broken or preserved \cite{Gaiotto:2014kfa,Bhardwaj:2023fca,Lootens:2024gfp}. In this framework, the correlation functions of Wilson loops serve as order parameters probing the (non-)spontaneous symmetry breaking of the center $1$-form symmetry that pure Yang-Mills exhibits \cite{Gaiotto:2014kfa}.

However, a similar criterion is not available for theories lacking such $1$-form symmetries. A key example of this is QCD, where dynamical matter in the fundamental representation explicitly breaks center symmetries that would otherwise be present. The absence of a Landau paradigm description reflects the lack of a sharp deconfinement phase transition for this class of theories, a phenomenon known as Higgs-confinement continuity \cite{Fradkin:1978dv,Banks:1979fi,Osterwalder:1977pc}. As such, (de)confinement is in general understood as a property of a subregion of the parameter space of a phase, rather than the phase itself.
However, for certain submanifolds of the parameter space, more refined diagnostics may distinguish these regimes as distinct phases, as shown e.g. in \cite{Dumitrescu:2023hbe}. 

The aim of this letter is to demonstrate that it is possible to differentiate the confined and the deconfined adjoint Higgs phases of scalar QCD at $T=0$ and fixed mass in terms of the baryon symmetry.
This is made possible by generalizing to higher dimensions the correspondence between the excitations in a gapped theory and representations of the strip algebra, which was developed in \cite{Cordova:2024vsq,Cordova:2024iti} for $(1+1)$D systems (see also \cite{Cordova:2024nux,Cordova:2024goh}).\footnote{For this technique to be applicable, both the $T=0$ and gapped requirements are important assumptions. This is why we discuss the adjoint Higgs phase instead of the more conventional Coulomb phase.} That is, we analyze the suitable higher representation theory that captures not only particle excitations, but also higher-dimensional excitations like strings in a specified phase of the theory. 
For the confined phase, for example, we will see that the representation theory correctly captures the existence of confining strings. In comparison, as expected we find no such behavior in the Higgs phase, where we instead detect center vortices around which bare quarks pick up an Aharanov-Bohm phase. More precisely, we suggest that whenever the theory is found in a deconfined phase admitting bare quarks, it is natural to expect the presence of center vortices.

While these phases may not strictly fit within the Landau paradigm, their spectra can still be described by mapping from regimes where a more controlled characterization applies. This relies on the strip algebra's ability to track how the matter content of a theory is re-organized with respect to symmetries exhibited across the RG flow. Beyond our case of interest, this work demonstrates that the higher representation theory of strip algebras provides a systematic framework for characterizing phases and excitations in a broader context. We aim to offer an extended analysis, including a more detailed study of scalar QCD, in the near future \cite{WIP}.\footnote{We have also been informed of the upcoming work \cite{Seiberg-Seifnashri} that has some overlap with our work.}

\section{Higher Strip Algebras} \label{sec:higher-strips}

\subsection{Review of \texorpdfstring{$(1+1)$}{(1+1)}-dimensions}

Before we discuss the higher representation theory of strip algebras in $(3+1)$D, we proceed to give a short review of the $(1+1)$D case following \cite{Cordova:2024iti}. See \cite{Kitaev:2011dxc,Lan:2013wia,Williamson:2017uzx,Bullivant:2019fmk,Bridgeman:2019wyu,Bridgeman:2022gdx,Copetti:2024dcz,Inamura:2024jke,Bhardwaj:2024igy,Choi:2024tri,Choi:2024wfm,JohnsonFreydReutter:Hopf,JohnsonFreydReutter:QuantumHpty} for other accounts, and Appendix \ref{sec:end-matter:cat-sym} for a brief review of categorical symmetries. We denote the strip algebra in both cases as $\str_\cC(\cM)$, where $\cC$ is a (higher) fusion category, and $\cM$ is a $\cC$-module (higher) category. $\cC$ is the symmetry of the theory under consideration and $\cM$ encodes the set of its spatial boundary conditions closed under the action by $\cC$.
Then, $\str_{\cC}(\cM)$ captures the action of $\mathcal{C}$ on the states of the theory placed on an interval, where the boundary conditions on the left and right boundaries are given by $\cM$. Via operator-state correspondence, this encapsulates how the symmetry acts both on local operator states and solitons of the theory.

The $(1+1)$D strip algebra $\str_\cC(\cM)$ is generated by elements representing the following figure:
\begin{figure}[h!]
\includegraphics{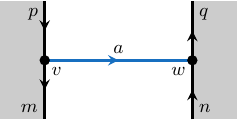}
\label{fig:2d-strip}
\end{figure}

\noindent Here $a\in \Obj(\cC)$ are the bulk topological lines, $m,\,n,\,p,\,q \in \Obj(\cM)$ are boundary lines, and $v\in \Hom_\cM(p,m),\,w\in \Hom_\cM(n,q)$ are topological junctions between boundary lines. The action of $\cC$ on the boundary conditions is then given by the fusion of bulk and boundary lines.

From a physical perspective, the choice of boundary conditions is dictated by the IR TQFT to which the theory flows, assuming it is in a \emph{gapped phase}. As a result, $\mathbf{Str}_\mathcal{C}(\mathcal{M})$ encodes how the UV symmetry $\mathcal{C}$ manifests in the IR regime.
In \cite{Cordova:2024iti,Bhardwaj:2024igy}, it is emphasized that the vacua and excitations --- particles and solitons --- are in the representations of $\str_\cC(\cM)$.
When the symmetry is just a group, i.e. $\cC=\vect(G)$, the module categories corresponding to symmetry preserving and breaking phases are 
\begin{subequations}
\begin{align}
    \mathcal{M}_{\mathrm{pres}} &=  \vect\, , \label{eq:module-pres}\\
    \mathcal{M}_{\mathrm{SSB}} &= \mathcal{C} = \vect(G) \, . \label{eq:module-SSB}
\end{align}
\end{subequations}
In particular, each of the module categories contains either 1 or $|G|$ simple objects, respectively, labeling the vacua in each phase.
For the former case, the strip algebra $\str_{\vect(G)}(\vect)\cong \bC[G]$ is the group algebra and its representation theory is equivalent to the ordinary group representation theory. For the latter case, the strip algebra $\str_{\vect(G)}(\vect(G))$ is a groupoid algebra encoding the non-linear realization of the $G$ symmetry.

Generally, the representation category of the strip algebra is given by the dual category:
\begin{equation} \label{eq:dual-cat}
    \rep[\str_\cC(\cM)] = \cC^*_\cM \coloneq \Hom_\cC(\cM,\cM).
\end{equation}
As we can scatter the excitations in $(1+1)$D, there should be tensor products in this representation category governing the \emph{selection rule} of such scatterings.
The tensor product is manifest in the right-hand-side of the formula \eqref{eq:dual-cat} and algebraically it comes from the weak Hopf algebra (WHA) structure of $\str_{\cC}(\cM)$. In the following, we assume that the same relation \eqref{eq:dual-cat} applies to higher dimensional theories to deduce the higher representation theory governing their excitations. Similarly, the representation category $\mathcal{C}^*_\mathcal{M}$ characterizes the organization of the IR operator content—encompassing local and extended operators, and solitons of various dimensions—under the action of the UV symmetry $\mathcal{C}$.

\subsection{Strip algebras in \texorpdfstring{$(d+1)$}{(d+1)}-dimensions}

\begin{figure}[t!]
\includegraphics{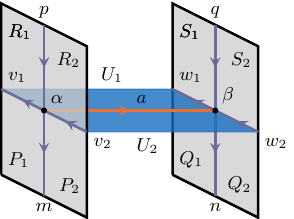}
\caption{A $1$-morphism $(a,p,q,n,m,\alpha,\beta)$ of the strip algebra between the objects $(U_i,P_i,Q_i,R_i,S_i,v_i,w_i)_{i=1,2}$.}
\label{fig:1homs-strip-algebra}
\end{figure}

The strip algebra in $(d+1)$-dimensions is a $d$-algebra, meaning it forms a $d$-category.
In our context, we take the strip algebra to act on flat space $\mathbb{R}^d$ with an explicit IR cutoff in one direction; that is the space is effectively the product of the interval $\mathcal{I}_L$ of a large length $L$ and the open space $\mathbb{R}^{d-1}$.
The choice of working with only a finite direction allows us to probe excitations with finite energy per unit volume. 
The objects of the strip algebra in the case of $d=2$ are generated by diagrams of the form
\begin{figure}[h!]
\includegraphics{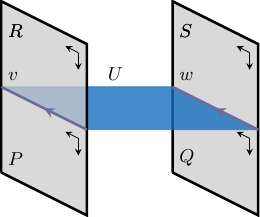}
\label{fig:3d-strip-objects}
\end{figure}

\noindent where $U \in \Obj(\cC)$, $P, Q, R, S \in \Obj(\cM)$ and $v,\,w\in 1\Hom(\cM)$. The 1-morphisms between two objects $(U_i,P_i,Q_i,R_i,S_i,v_i,w_i)_{i=1,2}$ are associated to an element $a\in 1\Hom(\cC)$, $p,\,q,\,m,\,n \in 1\Hom(\cM)$ and $\alpha,\,\beta\in 2\Hom(\cM)$, as depicted in FIG. \ref{fig:1homs-strip-algebra}. 
More precisely, $a$ is a 1-morphism in $\cC$ between $U_1$ and $U_2$, and, for example, $\beta$ is a 2-morphism in $\cM$ between $a\otimes(n\otimes w_1)$ and $w_2\otimes q$.
The higher strip algebra has a WHA structure like in $(1+1)$D, but we leave a full analysis of these details for future work. As in the $(1+1)$D case, we regard the module $\cM$ as specifying the IR gapped phase. In higher dimensions, a gapped phase can be topologically ordered, and the corresponding $\cM$ is non-trivial even when the vacuum degeneracy is trivial.
For example, for a finite abelian group $A$, the choice $\cM=d\vect(A^{[1]})$ represents the phase where the IR TQFT is the topological $A$-gauge theory. The case of interest to us is $d=3$, where the 1-morphism $a$ would be a topological surface associated to a finite 1-form symmetry $A^{[1]}$ with all other bulk operators trivial such that $\cC=3\vect(A^{[1]})$. If we pick $\cM = \cC = 3\vect(A^{[1]})$ then the boundary surfaces are labeling the holonomy in the IR $A$-gauge theory; the action of $a$ on these lines corresponds to shifting the holonomy. If we pick $\cM = 3\vect$ instead, then the boundary surfaces are trivial and $a$ will act only on Wilson lines inserted into the bulk, while the boundary surfaces remain trivial. In $(d+1)$-dimensions, it is expected that the representation category of the higher strip algebra $\mathcal{C}^*_\mathcal{M}$ is a $d$-fusion category, similarly to $(1+1)$D.\footnote{There has been progress towards this statement in the mathematics literature in the case of representation categories of 2-Hopf algebras \cite{Green:2023qqr}, though a proof for weak $d$-Hopf algebras has yet to be found for $d\geq 2$.} In this framework, the $p$-morphisms of $\cC_\cM^*$ correspond to representations of $(d-p)$-dimensional operators and solitons, thus providing an organized view of the symmetry action on the IR spectrum of a theory.

\section{Phases of Yang-Mills Theory} \label{sec:pure-ym}

We now wish to consider the structure of the strip algebra of $(3+1)$D pure Yang-Mills theory with gauge group $\mathcal{G}=SU(N)$ in both the confined and adjoint Higgs phases. We choose the spacetime to be $\bR^3\times \cI_L$ where we assume $L \gg L'$ where $L'$ is any other length scale in the theory. We let $t,x,y$ and $z$ be the coordinates of the spacetime, with $z$ being the direction of the interval $\cI_L$. 
With no additional matter, the theory exhibits a $1$-form symmetry $\bZ_N^{[1]}$ generated by Gukov-Witten operators associated to the center $Z(SU(N)) \cong \bZ_N$, which is preserved as the theory flows to the confined phase in the IR. 
Another possibility we consider is the addition of adjoint scalar matter, which Higgses the gauge group down to its center $\mathbb{Z}_N$. In this case, the corresponding $1$-form symmetry is spontaneously broken and the theory in the IR flows to the deconfined phase \cite{tHooft:1977nqb,Abrikosov:1956sx,Nielsen:1973cs} (see e.g. \cite{Komargodski:2017smk} for a higher-form symmetry perspective). In both instances, the symmetry we start with is $\bZ_N^{[1]}$, corresponding to the symmetry category $\mathcal{C}_\mathrm{YM}=3\mathbf{Vec}({\bZ^{[1]}_N})$, though it is realized differently in the IR in each phase.

Analogously to the $(1+1)$D cases, the module category of $\mathcal{C}_\mathrm{YM}$ for the symmetry preserving (confining) phase and the spontaneously broken (deconfined) phase are
\begin{subequations} \label{eq:module-choices}
\begin{align}
    \mathcal{M}_{\mathrm{conf}} &=  3\vect\, , \label{eq:module-conf}\\
    \mathcal{M}_{\mathrm{Higgs}} &= \mathcal{C} = 3\vect(\bZ^{[1]}_N) \, . \label{eq:module-higgs}
\end{align}
\end{subequations}
The corresponding strip algebras are higher versions of group and groupoid algebras, and its representation categories are ready to be computed using \eqref{eq:dual-cat}. The result in each phase is:
\begin{subequations} 
\begin{align}
    \rep_{\text{conf}} &= (3\vect(\bZ^{[1]}_N))^*_{3\vect} = 3\rep(\bZ^{[1]}_N) \label{eq:pure-conf-rep}, \\
    \rep_{\text{Higgs}} &= (3\vect(\bZ^{[1]}_N))^*_{3\vect(\bZ^{[1]}_N)} = 3\vect(\bZ^{[1]}_N) \, . \label{eq:pure-higgs-rep}
\end{align}
\end{subequations}
In particular, \eqref{eq:pure-conf-rep} follows from the fact that when the symmetry is realized properly in the IR so that the associated module $\cM_\mathrm{conf}$ is trivial, the representation theory is that of its (possibly extended) genuine operators, with no solitonic contributions \cite{Bartsch:2023pzl,Bhardwaj:2023wzd}. 
Instead, \eqref{eq:pure-higgs-rep} follows from the observation that in a fully spontaneously broken phase, bulk-boundary fusion coincides with the bulk fusion of defects. This is the higher analog of the familiar principle that symmetry actions permute the vacua of the theory in its SSB phase. Namely, we have the general fact that
\begin{equation}
    \cC \simeq \cC^*_\cC = \rep[\str_\cC(\cC)] \, ,
\end{equation}
known to hold in $(1+1)$D and assumed to remain valid in higher dimensions as well.

For the confined phase, where the symmetry is preserved, the strip-algebra representations coincide with the higher representation theory of the higher group $\mathbb{Z}_N^{[1]}$.
The representation category $\rep_{\text{conf}}$ is then generated by a non-trivial $1$-morphism, which should be identified with the confining string.
The symmetry operator $a_{p}[\Sigma]$ for $p \in \mathbb{Z}_N^{[1]}$ and a 2-surface $\Sigma$ detects a state $\ket{\text{c-string}(\gamma)}$ containing a confining string of unit charge along a line $\gamma$ via the intersection number $\Sigma\cdot \gamma$ in the space $\bR^2_{xy}\times \cI_L$:
\begin{equation}
    a_{p}[\Sigma] \ket{\text{c-string}(\gamma)} = e^{\frac{2\pi i q}{N} \Sigma\cdot \gamma} \ket{\text{c-string}(\gamma)} \, .
\end{equation}
The strip algebra includes $a_{p}[\Sigma]$ for $\Sigma$ spanning either $xz$ or $yz$ planes, and thus it can detect infinitely long confining strings stretching within the $\bR^2_{xy}$ plane.
Note that this is consistent with the linking action of the symmetry operator $a_p$ on the Wilson line that creates a confining string.

On the other hand, the representation category $\rep_{\text{Higgs}}$ is also generated by a non-trivial $1$-morphism, i.e.\ a string.
In this case, however, the string is a soliton associated to the spontaneously broken $\bZ_N^{[1]}$ symmetry.
The spontaneous breaking means that the holonomy of the gauge field along the infinitely long line within space can take any value in the center of the gauge group, $\bZ_N$. The string stretching along the $x$ direction in this phase connects two different holonomy values along $z$ or $y$ directions. The action of the symmetry operator $a_p[\Sigma]$ in this case shifts the holonomy values within $\Sigma$, while keeping the gauge field profile around the string unchanged.
Physically, the string soliton should be identified with the center vortex in the adjoint Higgs phase.

Mathematically, we have the equivalence between the representation categories of the two phases:
\begin{equation}
3\rep(\bZ_N^{[1]}) \simeq 3\vect(\bZ_N^{[1]})\, ,
\label{eq:equivalence-vect-rep}
\end{equation}
as both of them are generated by a string.
Physically, if we gauge the $\bZ_N^{[1]}$ symmetry, 
the two phases are exchanged -- the confined phase becomes the spontaneously broken phase of the magnetic $\bZ_N^{[1]}$ symmetry, and the Higgs phase is the phase preserving the magnetic symmetry. The equivalence \eqref{eq:equivalence-vect-rep} is a consequence of this duality.

\section{Scalar QCD} \label{sec:with-quarks}

The preceding section set the stage for our analysis of scalar QCD.
We extend our previous discussion by incorporating a massive scalar quark transforming in the fundamental representation of $SU(N)$, with an associated $U(1)_B$ baryonic symmetry.
The introduction of fundamental matter explicitly breaks the $\bZ_N^{[1]}$ $1$-form symmetry of pure Yang-Mills, as the Wilson lines can end on these local operators and are then screened to the trivial line. There exists also a non-trivial quotient between the gauge symmetry group and the baryonic symmetry, due to the possibility of reabsorbing a $\bZ_N \subset U(1)_B$ transformation via a gauge transformation. Namely, fundamental quarks $\phi^\alpha$ are invariant under the simultaneous transformations of these two symmetries. Therefore, the gauge and global symmetry groups combine to form the structure group
\begin{equation}
    \frac{SU(N) \times U(1)_B}{\bZ_N}\, .
    \label{eq:structure-group}
\end{equation}
While the quotient does not affect the global form of the global symmetry group, which is still $U(1)_B/\bZ_N \cong U(1)$, it does play an important role in how the symmetries are matched along the RG flow. To better capture this distinction, we focus on the behavior of a generic subgroup $\bZ_k \subset U(1)_B$, assuming $k=Nk_0$ for some $k_0 \geq 1$. Importantly, it is only the quotient $\bZ_{k_0} \cong \bZ_k / \bZ_N$ that is a global symmetry of the system. From a categorical perspective, the corresponding fusion $3$-category is then $\cC_\mathrm{QCD}=3\vect(\bZ_{k_0}^{[0]})$.

To analyze, it is convenient to assume that the quark mass $m_\phi \gg \Lambda_\mathrm{conf}$, where $\Lambda_\mathrm{conf}$ is the confining scale.
We can then look at the theory at an intermediate energy scale $\Lambda_{\mathrm{YM}}$, such that $m_\phi \gg \Lambda_\mathrm{YM} \gg \Lambda_{\mathrm{conf}}$, at which we can integrate out the massive quarks and recover pure Yang-Mills theory with its emergent $\bZ_N^{[1]}$ $1$-form symmetry.
For the Higgs phase, which corresponds to the choice of module $\cM_\mathrm{Higgs}=\cC_\mathrm{YM}$, we also assume the existence of adjoint matter neutral with respect to $U(1)_B$, such that to first order the Higgsing of the gauge group can be neglected at the scale $\Lambda_\mathrm{YM}$.

We can then define a symmetry homomorphism associated to the RG flow from scalar QCD to pure Yang-Mills as
\begin{equation}
    f:\mathcal{C}_{\mathrm{QCD}} \to \mathcal{C}_{\mathrm{YM}}\, .
    \label{eq:C-RG-flow}
\end{equation} 
Consider the short exact sequence describing the gauged diagonal $\bZ_N$ subgroup of $U(1)_B$, namely 
\begin{equation}
    \bZ_N\rightarrow \bZ_k \rightarrow \bZ_{k_0} \, .
\end{equation}
This yields the fiber sequence 
\begin{equation}
     \bZ_k \rightarrow \bZ_{k_0} \xrightarrow{f} B\bZ_N \, ,
    \label{eq:puppe-QCD}
\end{equation}
where $f:\bZ_{k_0} \rightarrow B\bZ_n$ is the Bockstein map. This is the RG flow homomorphism describing \eqref{eq:C-RG-flow}; we provide a proper justification of this in Appendix \ref{sec:end-matter:sym-hom}.

Strictly speaking, $f$ as in \eqref{eq:puppe-QCD} is a (higher) group homomorphism, but it can be promoted to a linear functor between the full symmetry categories 
\begin{equation}
    f: 3\vect(\bZ_{k_0}^{[0]}) \rightarrow 3\vect(\bZ_N^{[1]}).
\end{equation}
Naturally, this map will have an effect on the strip algebra, which we recall captures the organization of the IR TQFT data in terms of the symmetry defects present at some (possibly higher) energy scale. Given a choice of $\mathcal{C}_{\mathrm{YM}}$-module $\mathcal{M}$ from \eqref{eq:module-choices}, we can recover a corresponding $\mathcal{C}_{\mathrm{QCD}}$-module via pullback, $f^*\mathcal{M}$. This will describe the symmetry action on the IR TQFT boundary conditions in terms of the deep UV symmetry $\mathcal{C}_{\mathrm{QCD}}$ rather than $\mathcal{C}_{\mathrm{YM}}$, thus inducing a map between the strip algebras
\begin{equation}
\str_{\mathcal{C}_{\mathrm{QCD}}}(f^*\cM) \to     \str_{\mathcal{C}_\mathrm{YM}}(\cM)\, .
\end{equation}
Understanding the IR behavior with respect to the symmetry present at an intermediate scale -- in this case $\Lambda_{\mathrm{YM}}$ -- allows us to deduce how the same matter content reorganizes from a deeper UV scale, i.e. $m_\phi$.
In fact, by $f^*$ yielding a map between strip algebras, it provides a pullback map between the associated representation categories
\begin{subequations}
\begin{align}
    &\rep^{\mathrm{quark}}_{\mathrm{Higgs}} := f^*\rep_{\mathrm{Higgs}} \, , \label{eq:rep-q-higgs}\\
    &\rep^{\mathrm{quark}}_{\mathrm{conf}} := f^*\rep_{\mathrm{conf}} \, , \label{eq:rep-q-conf}
\end{align}
\end{subequations}
which thus describe how the deep IR TQFT excitations are organized from the perspective of $\mathcal{C}_\mathrm{QCD}$ defects.

To better understand how to compute \eqref{eq:rep-q-higgs}, we discuss first an easier setup. Consider the example of a theory with a global $0$-form abelian symmetry group $G^{[0]}$, such that in the IR the symmetry is spontaneously broken to a normal subgroup $H^{[0]} \subseteq G^{[0]}$. The vacua of the theory are then labeled by $G/H$.
Then, we have an exact sequence
\begin{equation}
    H \to G \to G/H \, ,
    \label{eq:G-H-extension}
\end{equation}
which describes $G$ as a central extension of $G/H$ by $H$, determined by a $2$-cocycle $\omega \in H^2(BG/H,H)$. The class $\omega$ appears in the exact sequence
\begin{equation}
\dots \to BG \to BG/H \overset{\omega}{\to} B^2 H \to  \dots \, ,
\end{equation}
and it can be seen as the invariant measuring the failure of $G/H$ defects to be lifted to $G$ defects. The $G^{[0]}$-module describing the boundary conditions in this setup is $\cM=d\vect((G/H)^{[0]})$, and so the associated representation category is 
\begin{equation}
    d\rep(H^{[0]})\boxtimes_\omega d\vect((G/H)^{[0]}) \, .
    \label{eq:G-H-rep}
\end{equation} 
Here $d\rep(H^{[0]})$ describes the representation of particle-states in terms of the IR-preserved symmetry $H^{[0]}$, and $d\vect((G/H)^{[0]})$ that of domain walls between the vacua, with the action of $G$-defects permuting their classes in the coset $G/H$. The twist $\omega$ captures the phase shift felt by the particle when it passes through a domain wall. Its presence can be inferred by interpreting the $d\vect((G/H)^{[0]})$ factor as the symmetry that would remain after gauging the subgroup $H^{[0]}$, in which case $\omega$ corresponds to the mixed anomaly with the emergent dual symmetry $\hat{H}^{[d-2]}$ \cite{Bhardwaj:2022dyt,Tachikawa:2017gyf,Cordova:2024iti}.

The construction of the representation category in \eqref{eq:G-H-rep} is functorial with respect to $\omega: BG/H \to B^2H$, implying that similar results apply to generic higher group maps. Specifically, in the case of scalar QCD in the Higgs phase, one can replace \eqref{eq:G-H-extension} with \eqref{eq:puppe-QCD}, where $\omega \in H^2(B^2 \bZ_N, \bZ_k)\cong \bZ_{N}$. Then, it follows that 
\begin{equation}
    \rep^\text{quark}_\text{Higgs} \cong 3\rep(\bZ_k^{[0]}) \boxtimes_\omega 3\vect(\bZ_N^{[1]}) \, .
\end{equation}
The first factor captures the charges of non-trivial classes of genuine operators via its 2-morphisms, which in this case correspond to particle states. Notice the full baryon symmetry $\bZ_k^{[0]}$ is revealed rather than its gauge-quotiented counterpart $\bZ_{k_0}^{[0]}$. This reflects the fact that the Higgsing of the gauge field exposes the bare quarks, making them visible in the Higgs phase. The second factor is the same as $\rep_\mathrm{Higgs}$ from the $\Lambda_\mathrm{YM}$ scale, which means that the UV symmetry $\bZ_{k_0}^{[0]}$ is still capable of probing IR center vortices through its flow to the $1$-form $\bZ_N^{[1]}$.
The appearance of the twist $\omega$ describes the Aharonov-Bohm phase a quark acquires when going around the vortices.
This is to be expected, as center vortices with unit $\bZ_N^{[1]}$ charge are characterized by fractional $1/N$ magnetic flux, reflecting the mixed 't Hooft anomaly between electric and magnetic symmetries of gauge theories.

The connection between the existence of bare quarks and center vortices just found is actually a general phenomenon, independent of the specific hierarchy of scales considered. Indeed, the preceding discussion suggests that if $\rep[\str_{\cC}(\cM)]$, for some module $\cM$, includes bare quarks, then there must exist a module functor $\mathcal{M} \to 3\vect(\bZ_N^{[1]})$. Consequently, $\rep[\str_{\mathcal{C}}(\mathcal{M})]$ must also include center vortices. This relationship will be explored in greater detail in the full paper \cite{WIP}.

Now consider the confined phase and its representation category \eqref{eq:rep-q-conf}. As previously argued, from the perspective of $\mathcal{C}_\mathrm{YM}$ the IR TQFT is described by $\mathcal{M}_\mathrm{conf} = 3\vect$. Due to its triviality, this module remains unchanged even after the pullback, namely
\begin{equation} 
f^* \mathcal{M}_\mathrm{conf} \cong 3\vect \,. 
\end{equation}
Therefore, it follows we must have 
\begin{equation}
    \rep_\text{conf}^\text{quark} =(3\vect(\bZ_{k_0}^{[0]}))^*_{3\vect}\cong 3\rep(\bZ_{k_0}^{[0]}).
    \label{eq:rep-quark-conf}
\end{equation}
At the scale $\Lambda_\mathrm{YM}$, the only non-trivial equivalence classes in $\rep_\mathrm{conf}$ were the $1$-morphisms, generated by the confining string. In contrast, the non-trivial irreducible classes in \eqref{eq:rep-quark-conf} occur at the level of $2$-morphisms and thus classify particles. Specifically, their detectable classes must carry charges only under $\bZ_{k_0}^{[0]}$, imposing the condition that their $U(1)_B$ charges satisfy $q = 0 \mod N$. The genuine particle content in the IR that is accessible through $\mathcal{C}_\mathrm{QCD}$ then consists solely of baryons, which demonstrates that bare quarks are excluded in this
phase of the theory, deserving the name ``confined phase''.
This also implies that the state $\ket{\text{c-string}(\gamma)}$, which contains an infinitely stretched confining string, is transparent with respect to $\mathcal{C}_\mathrm{QCD}$ and describes a condensation object in $3\rep(\bZ_{k_0}^{[0]})$. Consequently, the string must admit operators $\mathcal{Q}(x)$ at its endpoints, which, in turn, transform under a $\bZ_N$-projective representation of $\mathcal{C}_\mathrm{QCD}$. These operators are identified with quarks, which thus are condensed, with their projective $\bZ_{k_0}^{[0]}$-representation being identified by their corresponding $U(1)_B$-representation.
By further analyzing the pullback behavior of the original $\mathcal{C}_\mathrm{YM}$-representation of the confining string, we can not only deduce the existence of its endpoints and of baryons, but also directly probe the internal structure of the latter. For a detailed discussion of this, we refer to Appendix \ref{sec:end-matter:baryon-structure}.

\onecolumngrid
\acknowledgments
We would like to thank Mohamed Anber, Thomas Bartsch, Mathew Bullimore, Samson Chan, Mendel Nguyen, Jamie Pearson and Tin Sulejmanpasic for helpful discussions. We also thank Daniel Brennan, Clay Córdova, Nicholas Holfester and Sahand Seifnashri for useful comments on an earlier draft. We thank Aleksey Cherman and Masanori Hanada for helpful comments on the first version of this work. FG is funded by the STFC grant ST/Y509334/1. KO is supported by JSPS KAKENHI Grant-in-Aid No.22K13969 and No.24K00522. AG and KO also acknowledge support by the Simons Foundation Grant \#888984 (Simons Collaboration on Global Categorical Symmetries).
\twocolumngrid

\bibliography{refs}

\appendix

\section{Review of Categorical Symmetry} \label{sec:end-matter:cat-sym}

Throughout this letter, we describe discrete higher-form symmetries in terms of categories. We aim to review briefly the notation and conventions that we use, for more details we refer to \cite{Chang:2018iay,Bhardwaj:2017xup,Shao:2023gho,Bartsch:2023wvv, etingof2015tensor,douglas2018fusion2categoriesstatesuminvariant,Johnson_Freyd_2022,D_coppet_2022,decoppet2024classificationfusion2categories}. Suppose we have a $\bZ_N^{[p]}$ symmetry in a bosonic $(d+1)$-dimensional system. Formally, this means that the symmetry is described by the fusion $d$-category\footnote{The fusion $d$-category $d\vect$, of which $d\vect(\bZ^{[p]}_N)$ is a $\bZ_N^{[p]}$-enriched version, captures only oriented $d$-dimensional TQFTs that admit gapped boundaries. While in $d\ge 3$ this can impact scenarios like in the analysis of the chiral symmetry \cite{Cordova:2022ieu,Choi:2022jqy,Copetti:2023mcq}, it is not a relevant restriction for our discussion.} $\mathcal{C}=d\vect({\bZ^{[p]}_N})$. From a physical perspective, $k$-morphisms of this category correspond to $k$-form symmetries of the system, or in other words, its topological defects of codimension $k+1$. This means that, for example, $\cC=3\mathbf{Vec}(\bZ^{[1]}_N)$ has the following structure:
\begin{itemize}
    \item $\Obj(\cC)$ is the set of oriented topological $3$-dimensional defects of the theory, which in this case are all equivalent to the trivial one;
    \item $1\Hom(\cC)$ is the set of oriented topological surfaces of the theory, seen as interfaces between $3$-dimensional topological defects. Their equivalence classes correspond to elements $g\in \bZ_N$;
    \item $2\Hom(\cC),\, 3\Hom(\cC)$ are the set of topological lines and points, respectively, each equivalent to the trivial one in this case.
\end{itemize}
If we consider instead $\bZ_N^{[0]}$, then $\cC=3\vect(\bZ_N^{[0]})$ would have oriented topological 3-dimensional operators with equivalence classes labeled by $g\in\bZ_N$, with all $k$-morphisms trivial. These two symmetry categories will be the main examples we consider in this letter.

A $p$-form symmetry $A^{[p]}$ acts on $(\ge p)$-dimensional operators of the theory, and the category that captures their representations is given by $d\rep(A^{[p]})$ \cite{Bartsch:2023pzl,Bhardwaj:2023wzd}. Here $k$-morphisms correspond to the representations of $(d-k-1)$-dimensional objects with respect to $A^{[p]}$. See Table~\ref{tab:degree-dim} for the correspondence in $d=3$ case. In particular, this category will always have at least non-trivial indecomposable $(d-p-1)$-morphisms, labeled by $q\in \hat{A}:=\Hom(A,U(1))$, each corresponding to a different irreducible representation of a $p$-dimensional object charged under $A^{[p]}$, which action is defined by linking. Labels for indecomposable $(< d-p-1)$-morphisms depend instead on the choice of $d$ and $A^{[p]}$. For instance, there may be indecomposable, non-trivial, $k$-morphisms lying in the trivial equivalence class, namely admitting morphisms to the trivial $k$-representation, say $\pi:X \to 1$. These objects correspond to defects of the theory which admits boundaries transforming under projective representations with respect to $A^{[p]}$. For more discussion of this we refer the reader to \cite{Bartsch:2022ytj,Bhardwaj:2022yxj,Bartsch:2023pzl,Roumpedakis:2022aik,Gaiotto:2019xmp}.

\begin{table}[h]
    \begin{tabular}{c|c|c}
    Morphism degree $k$ & Dimension & Excitation    \\
    \hline
    0 (object)      & 2+1                       & domain wall    \\
    1               & 1+1                       & string / vortex \\
    2               & 0+1                       & particle      
    \end{tabular}
    \caption{The correspondence between morphism degree $k$ in a higher representation category, the dimension of the corresponding physical objects, and interpretation of the objects as excitations in $(3+1)$D. For $k\ge 1$, the interpretation is valid for the case when the morphism is among the lower degree identity morphisms, and a general morphism can be understood as a junction ($k=1$) or junction among junctions ($k=2$).}
    \label{tab:degree-dim}
\end{table}

\section{RG Flow Symmetry Homomorphism} \label{sec:end-matter:sym-hom}

In this Appendix we motivate why the Bockstein map
\begin{equation}
    f: \bZ_{k_0} \to B\bZ_N \label{eq-app:Bockstein}
\end{equation}
captures the symmetry homomorphism $\mathcal{C}_\mathrm{QCD} \to \mathcal{C}_\mathrm{YM}$ describing the RG flow of scalar QCD. For completeness, we consider the full baryonic symmetry $U(1)_B$, rather than its subgroup $\bZ_k$.

To verify this claim, consider a non-trivial flat bundle $P$ for the global symmetry group $U(1)_B/\bZ_N$ on a spacetime manifold $X$, which isomorphism classes are classified by $H^1(X,U(1)_B/\bZ_N) \cong H^1(X,U(1))$. Given a good cover $\{U_\alpha\}$ of $X$, the bundle can be described via \v{C}ech cohomology by its transition functions $g_{\alpha \beta}:U_\alpha \cap U_\beta \to U(1)$, satisfying
\begin{equation}
    g_{\alpha \beta} g_{\beta \gamma} g_{\gamma \alpha} = 1\, , 
    \label{eq:U(1)B-quotient-bundle}
\end{equation}
over $U_\alpha \cap U_\beta \cap U_\gamma \not = \varnothing.$
In order to have a correctly defined background field for the theory, $P$ should lift to a bundle of the whole structure group \eqref{eq:structure-group}. This means one needs to lift $g_{\alpha \beta}$ to functions $\tilde{g}_{\alpha\beta} : U_\alpha \cap U_\beta \to U(1)_B$. However, in general $\tilde{g}_{\alpha\beta}$ do not need to satisfy a cocycle condition, but rather we may have that
\begin{equation}
    \tilde{g}_{\alpha\beta} \tilde{g}_{\beta \gamma}\tilde{g}_{\gamma\alpha}=\exp(2\pi i \theta_{\alpha \beta \gamma}/N) \,, \quad \theta_{\alpha\beta\gamma}\in \bZ/N\bZ \, .
    \label{eq:lift-cocycle}
\end{equation}
Namely, there might be a failure in lifting $U(1)_B/\bZ_N$ defects to proper $U(1)_B$ defects. Furthermore, this failure is captured by a $2$-cocycle $\theta_{\alpha \beta \gamma} \in Z^2(X,\bZ_N)$, as it is easy to check that
\begin{equation}
    (\delta \theta)_{\alpha\beta\gamma\delta} = \theta_{\alpha \beta \gamma} - \theta_{\alpha \beta \delta} + \theta_{\alpha \gamma \delta} - \theta_{\beta \gamma \delta} = 0 \mod N \, ,
\end{equation}
for $U_\alpha \cap U_\beta \cap U_\gamma \cap U_\delta \not = \varnothing$.
The reason for this allowed failure is the presence of dynamical $SU(N)$ gauge fields, which can compensate for this twist by suitable shifts in the transition data $\{h_{\alpha\beta}\}$ describing the $SU(N)$-bundles. The net effect is that the entire configuration $(\tilde{g}_{\alpha\beta},h_{\alpha\beta})$ ensure the bundle $P$ can always be understood as a proper UV background of the full structure group. 

After integrating out the massive quarks, the original $U(1)_B$ baryon symmetry is no longer present. Nevertheless, the resulting $SU(N)$-bundles are still subject to nontrivial constraints, which are RG flow invariant. Such twists can now be understood as having turned on a non-trivial background for the emerging $1$-form symmetry, determined by a class in $H^2(X,\bZ_N)$. In other words, the original obstruction of $P$ to be lifted to a $U(1)_B$-bundle, rather than a bundle of the whole structure group, is what describes the constraining remnant of the UV symmetry in the IR. 
As it is possible to verify, the homomorphism
\begin{equation}
    f_*^X :H^1(X,U(1)) \to H^2(X,\bZ_N) \, ,
\end{equation}
constructed in this way is the operational definition of the Bockstein map in ordinary cohomology. The existence of the associated map \eqref{eq-app:Bockstein} then follows by Brown's representability theorem.

\section{Inner Structure of Baryons} \label{sec:end-matter:baryon-structure}

To better understand the internal structure of the baryons, it is useful to analyze non-trivial classes of (higher) morphisms in $\rep_\mathrm{conf}$ and their fate under the pullback $f^*$. If we neglect the extended surface operators of the theory, then representations of strings and local operators with respect to $\mathcal{C}_\mathrm{YM}$ correspond respectively to objects and $1$-morphisms of the $2$-category \cite{Bartsch:2023pzl}
\begin{equation}
    2\rep(\bZ_N^{[1]}) \cong \mathrm{End}_{3\rep(\bZ_N^{[1]})}(1) \, .
    \label{eq:rep-equivalence}
\end{equation}
Its simple objects are identified with the choice of irreducible $2$-representations of $\bZ_N^{[1]}$, that is functors $\mu:B^2 \bZ_N \to 2\vect$.
Due to the absence of $0$-form components in $\mathcal{C}_\mathrm{YM}$, these maps must factor through the subgroupoid $\mathrm{Aut}_{2\vect}(1) \cong B^2 U(1)$, so that the equivalence classes are identified by 
\begin{equation}
    \pi_0 \text{Map}(B^2\bZ_N,B^2U(1)) \cong H^2(B^2\bZ_N,U(1)) \cong \bZ_N. 
    \label{eq:H2-strings}
\end{equation}
Each class $[\mu_q]$ of \eqref{eq:H2-strings} identifies the representation of a Wilson loop of corresponding $N$-ality $q$. Similarly, equivalence classes of $1$-morphisms correspond to classes of natural transformations between such functors. By analogous arguments, these are classes of maps $\pi_0 \text{Map}(S^1,\text{Map}(B^2 \bZ_N, B^2 U(1))) \cong H^2(B^3 \bZ_N,U(1))$, which in turn vanish. Namely, as already claimed, no genuine particle excitation of the theory can be probed by $\mathcal{C}_\mathrm{YM}$.

Instead, from the perspective of $\mathcal{C}_\mathrm{QCD}$, the matter content organizes itself into $2\rep(\bZ_{k_0}^{[0]})$. Now classes of objects that are pulled backed from $2\rep(\bZ_{N}^{[1]})$ are given by maps $\mu_q \circ f: B\bZ_{k_0} \to B^2U(1)$, which are identified with the vanishing group $H^2(B\bZ_{k_0}, U(1))$.
It follows that as each $\mu_q \circ f$ is in the trivial class, they must correspond to condensation defects in $2\rep(\bZ_{k_0}^{[0]})$. This means the associated Wilson loops can end on local operators charged under $\mathcal{C}_\mathrm{QCD}$. These are quarks of charge $q$, that are described as $1$-morphisms $\mathcal{Q}_q:\mu_q \circ f \to 1$. 

Notice, however, that non-trivial classes of genuine particles correspond only to $H^3(B\bZ_{k_0},U(1))\cong \bZ_{k_0}$, so it is not possible to see these quarks isolated, but only baryons. 
Nevertheless, it becomes possible to understand the structure of the latter. Consider for instance the stacking of $N$ confined strings $W^{\otimes N}_1$. From the point of view of $\mathcal{C}_\mathrm{YM}$, this is a completely transparent line. This does not mean the corresponding object $\mu^N_1 \in 2\rep(\bZ_N^{[1]})$ is the identity line $1$, but rather that it is isomorphic to it, so that there are non-trivial $1$-morphisms $\mathcal{J}:1 \to \mu^{N}_1$, corresponding to operators on which $W^{\otimes N}_1$ can end. By virtue of $[\mu^N_1] = [1]$, their equivalence classes must correspond to those of genuine particles, which, from the perspective of $\mathcal{C}_\mathrm{YM}$, are absent. Such correspondence holds also when organizing the matter representations with respect to $\mathcal{C}_\mathrm{QCD}$, as we still have $[\mu^N_1 \circ f] = [1]$. This implies that for each non-trivial class of baryons, seen as maps $\mathcal{B}_p:1 \to 1$, there exists a corresponding non-trivial class of morphisms $\mathcal{J}_p: 1 \to \mu^{N}_1 \circ f$. Here, $p$ denotes the charge of a baryonic particle under $\bZ_{k_0}^{[0]}$, which is equivalently expressed as a charge of $pN$ under $U(1)_B$.
As we can verify, this equivalence map is provided precisely by the existence of quarks argued earlier, which allow to describe $\mathcal{B}_p$ as the composition
\begin{equation}
    \mathcal{B}_p \cong \mathcal{Q}_1^N \circ  \mathcal{J}_p \, .
\end{equation}
In other words, this makes precise the idea of a baryon $\mathcal{B}_p$ being a bound state of quarks linked by flux tubes through a $p$-valent junction, as depicted in FIG. \ref{fig:baryon-structure} below.
\begin{figure}[h!]
\includegraphics{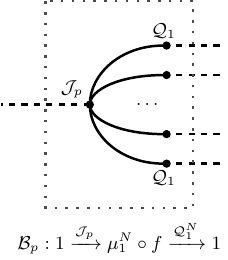}
\caption{Inner structure of a baryon with its $p$-valent junction.}
\label{fig:baryon-structure}
\end{figure}

\end{document}